\documentclass[12pt]{iopart}
\usepackage{epsfig}
\usepackage{iopams}

\begin{document}

\title[Dynamical instabilities of a resonator]{Dynamical instabilities of a resonator driven by a superconducting single-electron
transistor}%
\author{D.A. Rodrigues\dag, J. Imbers, T.J. Harvey and A.D. Armour}%
\address{School of Physics and Astronomy, University of
Nottingham, Nottingham NG7 2RD, United Kingdom}%

\ead{\mailto{\dag denzil.rodrigues@nottingham.ac.uk}}

% ----------------------------------------------------------------
\begin{abstract}
We investigate the dynamical instabilities of a resonator coupled to
a superconducting single-electron transistor (SSET) tuned to the
Josephson quasiparticle (JQP) resonance. Starting from the quantum
master equation of the system, we use a standard semiclassical
approximation to derive a closed set of mean field equations which
describe the average dynamics of the resonator and SSET charge.
Using amplitude and phase coordinates for the resonator and assuming
that the amplitude changes much more slowly than the phase, we
explore the instabilities which arise in the resonator dynamics as a
function of coupling to the SSET, detuning from the JQP resonance
and the resonator frequency. We find that the locations (in
parameter space) and sizes of the limit cycle states predicted by
the mean field equations agree well with numerical solutions of the
full master equation for sufficiently weak SSET-resonator coupling.
The mean field equations also give a good qualitative description of
the set of dynamical transitions in the resonator state that occur
as the coupling is progressively increased.

\end{abstract}
\submitto{NJP}

\maketitle
% ----------------------------------------------------------------

\section{Introduction}

Nanoelectromechanical systems in which a high frequency mechanical
resonator is coupled to a mesoscopic conductor\,\cite{Blencowe} have
been predicted to display a wide variety of different dynamical
behaviours depending on the nature of the conductor. When a
mechanical resonator is linearly coupled to the transport electrons
in either a quantum point contact\,\cite{MM,CG,Swede}, or a normal
state single-electron transistor\,\cite{mmh,ABZ,SET-expt}, the
electrons act on the resonator like an effective thermal
bath\,\cite{MM,mmh,ABZ,CG,Clerk,Swede}. Under such circumstances,
the resonator is damped and reaches an almost Gaussian steady state
whose width is set by the fluctuations in the motion of the charges
through the conductor. In contrast, where the electro-mechanical
coupling is non-linear\,\cite{shuttle,Usmani}, or the conductor is
close to a transport resonance\,\cite{SSET1,ria,bc}, the mechanical
resonator can be driven by the electrons into states of
self-sustaining oscillation.

In this paper we analyze the instabilities that arise in the
dynamics of a mechanical resonator coupled to a superconducting
single-electron transistor\,\cite{SSET1,SSET2,SET-expt2,SSET-expt}
(SSET) operated in the vicinity of a transport resonance. In the
SSET, transport resonances occur when states of the SSET island
differing by one Cooper pair are degenerate so that coherent Cooper
pair tunnelling between the island and one of the leads is possible.
The simplest such resonance, which we concentrate on here, is called
the Josephson quasiparticle (JQP)
resonance\,\cite{Choi,CSSET,SSET1,SSET2} and involves a cycle of
processes in which current flows via a combination of coherent
Josephson tunnelling between the SSET island and one of the leads,
followed by incoherent quasiparticle decays into the other junction.

When a SSET tuned to the vicinity of the JQP resonance is coupled to
a mechanical resonator, the resonator dynamics is very sensitive to
the precise choice of bias point for the SSET\,\cite{SSET1,SSET2}.
For operating points on one side of the JQP resonance, the SSET
damps the resonator and can be regarded as an effective thermal
bath, behaviour which was confirmed in recent
experiments\,\cite{SSET-expt}. In contrast, for operating points on
the other side of the resonance the electrical degrees of freedom
pump the resonator leading to the possibility of states of
self-sustaining oscillation. In this regime the resonator dynamics
can be investigated either by numerical solution of the quantum
master equation\,\cite{ria}, or provided the resonator is much
slower than the electrical degrees of freedom (as was the case in
recent experiments\,\cite{SET-expt,SSET-expt}), an effective
Fokker-Planck equation for the resonator can be
derived\,\cite{SSET1,bc}.

Numerical solution of the SSET-resonator master equation\,\cite{ria}
revealed interesting similarities with a quantum optical device
known as the micromaser. In a
micromaser\,\cite{micromaser,micromaser_review}, a cavity resonator
is driven by the passage of a steady stream of excited two-level
atoms. As the atom-cavity interaction is increased, the resonator
undergoes a sequence of dynamical transitions leading in some cases
to non-classical steady states. A corresponding set of dynamical
transitions was found to occur in the SSET-resonator system together
with regions of non-classicality.

Here we use an alternative approach, namely a mean field description
to analyze the instabilities in the dynamics of the SSET-resonator
system. This kind of approach has been used extensively in the
analysis of non-linear quantum optical systems where its usefulness
has been well established\,\cite{WM,mabuchi}. The mean field
equations we derive provide a relatively compact description of the
system and their stability properties are readily analyzed using the
techniques of classical dynamical systems theory. Although the mean
field description neglects some of the correlations in the system,
comparison with numerical results from the full master equation show
that for sufficiently weak coupling the mean field theory is close
to being quantitatively correct. Although quantitative agreement is
poor at stronger couplings, the mean field equations still give a
good qualitative description of the system's dynamics displaying a
similar sequence of transitions to that found
numerically\,\cite{ria}.

Although we will use language appropriate to a nanomechanical
resonator throughout this paper, we anticipate that much of our
analysis will also be relevant to the case of a superconducting
resonator coupled to a SSET. Photon assisted satellites of JQP peaks
have been observed in experiments in which a SSET was coupled to
microstrip transmission line\,\cite{rimberg}. Furthermore, recent
experiments have demonstrated coherent coupling between a
superconducting stripline resonator and a Cooper-pair box with a
coupling Hamiltonian between the resonator and the charges on the
box very similar to the mechanical case\,\cite{SCR}. Such systems
would only need to be modified to allow quasiparticle tunnelling off
the Cooper-pair box\,\cite{Nakamura} to become analogous to the
device considered here and hence it seems likely that a range of
dynamical instabilities similar to that which we find for the
mechanical case could also be produced in a superconducting
resonator.

The outline of this paper is as follows. In Sec.\ 2 we introduce our
model of the SSET-resonator system and give the appropriate quantum
mechanical master equation.  The mean field equations are derived in
Sec.\ 3. In Sec.\ 4 we transform the mean field equations into plane
polar coordinates and exploit the fact that the amplitude of the
resonator oscillations is slowly changing to derive an effective
amplitude dependent damping of the resonator due to the SSET. We
then show that this quantity can be used to predict the presence of
limit cycles in the resonator dynamics. We compare the limit cycle
solutions predicted by the mean field theory with numerical
calculations using the full master equation in Sec.\ 5. Then in
Sec.\ 6 we briefly consider the implications of our theoretical
calculations for experiments on nanomechanical-SSET systems.
Finally, in Sec.\ 7 we draw our conclusions. In the Appendixes we
give additional details about the derivation of the master equation
for the SSET-resonator system and the stability analysis of the mean
field equations.

\section{Master Equation}\label{sec:me}

\begin{figure}[t]\centering{
 \epsfig{file=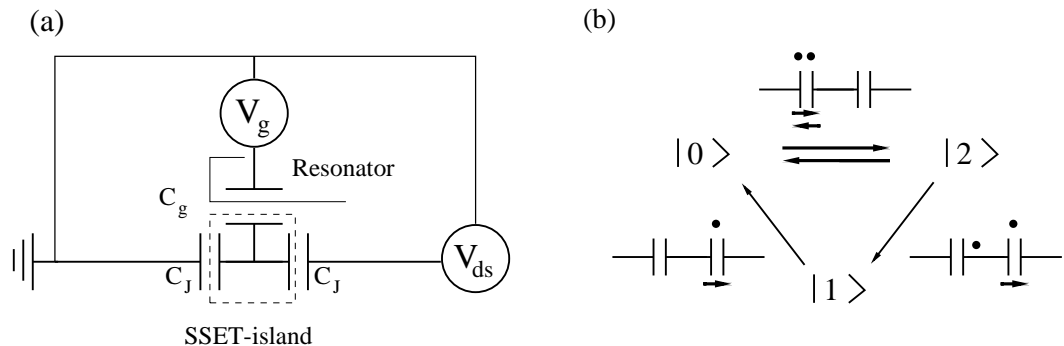, width=14.0cm}
} \caption{(a) Circuit diagram for the coupled SET--resonator
system. Under the bias voltage shown the electrons flow through the
SSET island from left to right. (b) Summary of charge processes
involved in the JQP cycle.} \label{fig:schema}
\end{figure}

The SSET-resonator system we consider is shown schematically in
figure \ref{fig:schema}a. The SSET consists of an island linked to
leads by tunnel junctions with resistances $R_J$ and capacitances
$C_J$ which are taken to be equal for simplicity. The mechanical
resonator is treated as a single-mode harmonic oscillator, with
frequency $\omega$, which forms a movable gate capacitor. The
equilibrium position of the resonator is a distance $d$ from the
SSET island and we assume that the displacement of the resonator
with respect to this equilibrium position, $x$, is always small in
comparison (i.e.\ $|x|\ll d)$. Under these circumstances the gate
capacitance can be expanded up to just linear order and we have
$C_g(x)=C_g(1-x/d)$, where $C_g$ is the capacitance at $x=0$.

The JQP resonance\,\cite{Choi,CSSET,SSET1,SSET2} which we are
interested in here occurs when two conditions are met. Firstly, at
the centre of the resonance there is no change in the energy of the
system when a Cooper pair tunnels between one of the leads and the
SSET island. Secondly, the bias voltage, $V_{ds}$, must be large
enough to allow quasiparticle decay processes between the island and
the other lead to occur. The charge processes involved in the JQP
resonance we consider are summarized in figure \ref{fig:schema}b.
Josephson tunnelling across the left  junction leads to coherent
oscillations between the SSET island states $|0\rangle$ and
$|2\rangle$ which differ by a single Cooper pair. These oscillations
are interrupted by the tunnelling of a quasiparticle from the island
into the right  lead which takes the island into charge state
$|1\rangle$. A further quasiparticle then tunnels from the island to
the right  lead, returning the island to state $|0\rangle$ and the
cycle begins again. The large electrostatic charging energy of the
small SSET island and the carefully controlled bias voltage ensure
that at low temperatures all other charge processes are strongly
suppressed.

The master equation for the SSET island charge and resonator is
obtained from the full Hamiltonian of the system by tracing out the
quasiparticle degrees of freedom\,\cite{Choi}. The steps involved in
this derivation (together with details of the approximations and
simplifications involved) are sketched in Appendix A. The final
result is a master equation of the form\,\cite{ria}
\begin{equation}
\dot{\rho}={\mathcal L}\rho=-\frac{i}{\hbar}[H_{\rm
co},\rho]+\mathcal{L}_{leads}\rho+\mathcal{L}_{damping}\rho.
\label{master}
\end{equation}
The evolution consists of a coherent part, described by the
Hamiltonian, $H_{\rm co}$, together with two dissipative terms
$\mathcal{L}_{leads}$ and $\mathcal{L}_{damping}$ which describe
quasiparticle decay from the island and the surroundings of the
resonator respectively. The effective Hamiltonian is given
by\,\cite{ria}
\begin{eqnarray}
H_{\rm co}&=&\Delta E |2\rangle \langle 2|
-\frac{E_J}{2}\left(|0\rangle\langle 2|+|2\rangle\langle 0|\right)
\\&&+\frac{p^2}{2m}+\frac{1}{2}m\omega^2x^2+ m\omega^2x_sx\left(|1\rangle\langle 1|+2|2\rangle\langle
2|\right),  \nonumber \label{Ham}
\end{eqnarray}
where $\Delta E$ is the energy difference between states $|2\rangle$
and $|0\rangle$, $E_J$ is the Josephson energy of the
superconductor, $x$ and $p$ are the canonical position and momentum
operators of the resonator and $m$ is the effective resonator mass.
The resonator-SSET coupling is described by the length-scale $x_s$
which measures the shift in the equilibrium position of the
resonator brought about by adding a single electronic charge to the
SSET island\,\cite{SSET2}. The tunnelling of quasiparticles from the
island is described by
\begin{eqnarray}
\mathcal{L}_{leads}\rho&=&-\frac{\Gamma}{2}\left[\left\{|2\rangle\langle
2| +|1\rangle\langle
1|,\rho\right\}_+\right.\\&&\left.-2\left(|1\rangle\langle
2|+|0\rangle\langle 1|\right)\rho\left(|2\rangle\langle
1|+|1\rangle\langle 0|\right)\right], \nonumber
\end{eqnarray}
where $\Gamma$ is the quasiparticle decay rate. Note that this is a
simplified expression which uses a single rate for the two decay
processes and neglects both the position dependence of the
quasiparticle rates and their dependence on bias point (these
approximations are discussed in more detail in Appendix A).
Simplified in this way, the master equation is also essentially
equivalent to that used to describe a resonator coupled to a double
quantum dot\,\cite{BL}.

The term which describes the effect of the resonator's surroundings
is
\begin{equation}
\mathcal{L}_{damping}\rho=-\frac{i\gamma_{ext}}{2\hbar}[x,\left\{p,\rho\right\}_+]
-\frac{\gamma_{ext}m\omega}{\hbar}(\overline{n}+1/2)[x,[x,\rho]],
\end{equation}
where $\gamma_{ext}$ represents the external damping and
$\overline{n}$ the equilibrium average resonator occupation that
would occur in the absence of the SSET. We use this form of the
oscillator damping kernel as opposed to the quantum optical form
which we have used elsewhere\,\cite{ria} as although it is less
convenient for numerical calculations, it leads to the correct
(translationally invariant) classical limit, which is essential if
we wish to derive appropriate mean field
equations\,\cite{Flindt,Kohen}.

The behaviour of the resonator depends very sensitively on the sign
of $\Delta E$. For $\Delta E>0$, Cooper pairs tend to absorb energy
from the resonator damping its
motion\,\cite{footnote_3,SSET1,SSET2}. In contrast, for $\Delta E<0$
and when the resonator is slow (i.e.\ $\omega\ll\Gamma$) the
resonator tends to absorb energy from the Cooper pairs and it is in
this regime that instabilities can occur. For faster resonator
speeds\,\cite{ria} ($\omega/\Gamma\ge 1$), absorbtion of energy by
the resonator from the SSET, and hence the location of
instabilities, becomes concentrated around points where $\Delta
E\simeq -j\hbar \omega$ with $j$ an integer.

\section{Mean Field Equations}

The mean field equations for the SSET-resonator system consist of
the set of equations of motion for the expectation values of all the
relevant SSET and resonator operators. They are derived by
multiplying the master equation by each operator in turn and taking
the trace:
\begin{eqnarray}
\dot{\langle{x}\rangle}&=& \langle v\rangle\\
\dot{\langle{v}\rangle}&=&-\omega^2(x+x_s[\langle p_{11}\rangle
+2\langle p_{22}\rangle])-\gamma_{ext}\langle v\rangle\\
\dot{\langle{p_{00}}\rangle}&=&\Gamma\langle p_{11}\rangle +i\frac{E_J}{2\hbar}(\langle\rho_{20}\rangle-\langle\rho_{02}\rangle)\\
\dot{\langle{p_{11}}\rangle}&=&\Gamma(\langle p_{22}\rangle-\langle p_{11}\rangle)\\
\dot{\langle{p_{22}}\rangle}&=&-\Gamma\langle p_{22}\rangle -i\frac{E_J}{2\hbar}(\langle\rho_{20}\rangle-\langle\rho_{02}\rangle)\\
\dot{\langle{p_{02}}\rangle}&=&i\frac{E_J}{2\hbar}(\langle
p_{22}\rangle-\langle p_{00}\rangle)-\frac{1}{2}\Gamma\langle
\rho_{02}\rangle+\frac{i}{\hbar}
 (\Delta E\langle \rho_{02}\rangle+2m\omega^2x_s\langle
 x\rho_{02}\rangle)
\end{eqnarray}
where  the SSET charge operators are defined as
$p_{ij}=|i\rangle\langle j|$ with $i,j=0,1,2$ and the expectation
values are defined by $\langle \dots \rangle={\rm Tr [\rho \dots]}$.
Note that, as discussed in Appendix A, the $p_{12},p_{10}$
components of the density operator decouple from the rest and can
safely be neglected from the mean field equations (and also the
master equation itself) as they do not affect the resonator.

In contrast to the simpler case of a resonator coupled to a normal
state SET\,\cite{ABZ,dar}, the equations of motion of the first
moments do not form a closed set. In other words, the dynamics of
the first moments depend on the behaviour of higher moments leading
to an infinite hierarchy of equations of motion for progressively
higher order moments of the system operators. In order to derive a
simple set of dynamical equations we need to make a semiclassical
approximation whereby the expectation value of a product of two
operators is replaced by the corresponding product of the
expectation values of the individual operators\,\cite{WM,mabuchi}.
Thus in this case we substitute $\langle x\rangle\langle
p_{02}\rangle$
 for $\langle xp_{02}\rangle$. This approximation
cannot be justified rigorously. Indeed, dropping the correlations
between $x$ and $\rho_{02}$ means that we lose the ability to
describe the noise in the system (which is determined by the
behaviour of the higher moments). Nevertheless, the semiclassical
approximation is well-known as a way of deriving a set of dynamical
equations which typically capture many of the important elements of
the dynamics of the corresponding quantum system\,\cite{mabuchi}. In
this case, we find that the much simpler mean field equations which
result from the semiclassical approximation provide a very useful
qualitative and, for sufficiently weak SSET-resonator coupling,
quantitative description of the different dynamical transitions
which the resonator can undergo.

Thus, having made the semiclassical approximation, and using the
conservation of probability ($\langle p_{00}\rangle +\langle
p_{11}\rangle + \langle p_{22}\rangle=1$) to eliminate one of the
charge variables, we obtain the following closed set of mean field
equations
\begin{eqnarray} \label{mfxv}
\dot{x}&=& v \label{mf1} \\
\dot{v} &=& -\omega^2 \left(x +x_s[ p_{11}+2p_{22}]\right) -\gamma_{ext} v\\
\dot{\alpha} &=& - \frac{1}{\hbar}\left(\Delta E + \hbar\omega\frac{x_sx}{x_q^2}\right) \beta - \frac{\Gamma}{2}\alpha \\
\dot{\beta} &=& \frac{1}{\hbar}\left(\Delta E +\hbar\omega
\frac{x_sx}{x_q^2}\right) \alpha - \frac{\Gamma}{2}\beta+\frac{
E_J}{2\hbar} (2p_{22}+p_{11}-1)\\
\dot{p}_{11} &=& \Gamma (p_{22} -  p_{11}) \\
\dot{p}_{22} &=& -\frac{E_J}{\hbar} \beta -\Gamma p_{22}
\label{mf5},
\end{eqnarray}
where $x_q^2=\hbar/(2m\omega)$ and we have dropped the angled
brackets for convenience. The quantities $\alpha$ and $\beta$ are
defined as the real and imaginary parts of $\langle p_{02}\rangle$
respectively.

Despite the decoupling of the second moment in the above equations,
the mean field equations reproduce many of the features found in the
dynamics of the full master equation. The time evolution of these
equations reveals the resonator relaxing to a fixed point for some
values of the parameters. For other values, the oscillations grow at
first, before settling into a limit cycle, i.e.\ an oscillation at
fixed amplitude. The  mean field equations also show clear evidence
of bistability for certain parameter values: in these cases the
long-time behaviour of the resonator depends on the choice of
initial conditions.

The fixed point solution of the mean field equations is obtained by
setting the time derivatives to zero in equations
(\ref{mf1})-(\ref{mf5}). As we discuss in Appendix B, the stability
of the fixed point can be established using standard
techniques\,\cite{mabuchi}.

\section{Analysis in radial coordinates}\label{sec:radial}

Although a straightforward stability analysis of the mean field
equations can tell us quite a lot about the dynamics of the system,
we find that transforming the mean field equations into plane-polar
coordinates and making one further simplifying assumption allows us
to proceed much further. The assumption we make is that the
SSET-resonator coupling and external damping are sufficiently weak
that the resonator's energy changes much more slowly than either its
phase or the charge state of the SSET, conditions which are readily
met for most practical implementations of the SSET-resonator system.
This type of approximation has already proved useful in describing a
variety of nanoelectromechanical and optomechanical
systems\,\cite{Usmani,SSET2,MHG}. In terms of our analysis of the
mean field equations, the assumption of a wide separation of time
scales between the evolution of the amplitude and phase of the
resonator allows us to derive an effective equation of motion for
the resonator amplitude from which we can determine the number and
location of stable limit cycle solutions\,\cite{MHG}.

We proceed by rewriting the mean field equations [equations
(\ref{mf1})-(\ref{mf5})] in plane polar co-ordinates $(A,\phi)$
which describe the amplitude and phase of the resonator, defined
through the relations $x-x_{fp}=A \cos \phi$ and $v=\omega A \sin
\phi$, where $x_{fp}$ is the fixed point resonator position. In
terms of $(A,\phi)$ the mean field equations take the form,
\begin{eqnarray}
\dot{A} &=&  - \gamma_{ext} A \sin^2 \phi-
\omega[x_{fp}+x_s(p_{11}+2p_{22})]\sin\phi \label{mfrphi1}\\
\dot{\phi} &=& -\omega - \gamma_{ext} \sin \phi \cos \phi-\frac{\omega}{A} [x_{fp}+x_s(p_{11}+2p_{22})]\cos \phi  \label{mfrphi2} \\
\dot{p}_{11} &=& \Gamma \left(p_{22} - p_{11}\right)  \label{mfrphi3}\\
\dot{p}_{22} &=&  \frac{E_J}{2\hbar} \beta -\Gamma p_{22} \label{mfrphi4}\\
\dot{\alpha} &=& - \left[\frac{\Delta E}{\hbar} +\omega x_s\frac{(x_{fp}+A\cos \phi)}{x_q^2}\right] \beta
 - \frac{\Gamma}{2}\alpha \label{mfrphi5} \\
\dot{\beta} &=&  \left[\frac{\Delta E}{\hbar} +\omega
x_s\frac{(x_{fp}+A\cos \phi)}{x_q^2}\right] \alpha -
\frac{\Gamma}{2}\beta+\frac{ E_J}{2\hbar} (2p_{22}+p_{11}-1)
\label{mfrphi6}.
\end{eqnarray}
Note that $x_{fp}$ is equal to $-x_s(p_{11}+2p_{22})$ when $p_{11}$
and $p_{22}$ have their fixed point values.

We now assume that the evolution of the resonator amplitude, $A$, is
sufficiently slow that during a {\it single} period of oscillation
it can be treated as a constant and that the phase evolution over
the same time can be approximated by  $\phi =\omega t$. The dynamics
of the electronic degrees of freedom can then be obtained for this
constant amplitude and steadily evolving phase. The resulting forced
dynamics of the SSET charge variables are then averaged over the
resonator period to calculate their effect on the resonator
amplitude $A$ which we characterize by an amplitude dependent
damping term\,\cite{Usmani,SSET1,MHG}, $\gamma_{SSET}(A)$. This
slow-$A$ approximation will certainly be appropriate in the vicinity
of a limit cycle solution of the mean field equations and will be
valid more generally so long as the free (uncoupled) evolution of
both the resonator and charge degrees of freedom is much faster than
the rate of change of the resonator amplitude in the coupled system,
i.e.\ for $\omega, \Gamma\gg \gamma_{ext}$ and sufficiently weak
SSET-resonator coupling. Whilst there is no simple way of evaluating
the strength of the SSET resonator coupling at which this
approximation breaks down, from equations (\ref{mfrphi1}) and
(\ref{mfrphi2}), we expect that this approach will be valid for a
given amplitude, $A$, if $x_s\ll A$ (note that if the populations
$p_{11}$ and $p_{22}$ remain much less than unity then the
conditions on $x_s$ will be less restrictive). Furthermore, it is
clear that in order to be consistent with these assumptions we
should obtain an effective damping $\gamma_{SSET}(A)$ whose
magnitude is much less than $\omega$ and $\Gamma$.

\subsection{Solving the electronic dynamics} \label{sec:fourier}

With a fixed amplitude, $A$, and a harmonically oscillating phase,
$\phi=\omega t$, the electronic degrees of freedom form a set of
four coupled differential equations with time dependent
coefficients. The dynamics of even this simplified system is still
non-trivial, but we can make progress by making use of our
assumption that the electronic degrees of freedom relax rapidly
compared to the resonator amplitude. We assume that transients in
the charge dynamics can be neglected and hence that the effect on
the amplitude of the resonator is dominated by the periodic response
of the charge degrees of freedom to the harmonic drive.

To extract the relevant periodic solutions for the charges we
rewrite the equations of motion in terms of a Fourier series
consisting of harmonics of the resonator frequency, defined by,
e.g.\
\[
p_{11}(t)=\sum_{n=-\infty}^{+\infty}p^n_{11}{\rm e}^{in\omega t}.
\]
The resulting equations for the Fourier coefficients of the
electronic variables are,
\begin{eqnarray}
i \omega n p_{11}^n &=& \Gamma \left(p_{22}^n-p_{11}^n\right) \label{eq:fc1}\\
i \omega n p_{22}^n &=& -\Gamma p_{22}^n- \frac{E_J}{2\hbar} \beta^n \label{eq:fc2}\\
i \omega n \alpha^n &=& - \left(\frac{\Delta E}{\hbar}+\omega
\frac{x_{fp}x_s}{x_q^2}\right) \beta^n- \omega \frac{Ax_s}{2x_q^2} (\beta^{n+1}+\beta^{n-1}) -\frac{\Gamma}{2}\alpha^n\label{eq:fc3}\\
i \omega n \beta^n &=&  \left(\frac{\Delta E}{\hbar}+\omega
\frac{x_{fp}x_s}{x_q^2}\right) \alpha^n
+\frac{E_J}{2\hbar}(2p^n_{22}+p^n_{11}-\delta_{n,0})\nonumber\\&& +
 \omega \frac{Ax_s}{2x_q^2} (\alpha^{n+1}+\alpha^{n-1})
-\frac{\Gamma}{2}\beta^n.\label{eq:fc4}
\end{eqnarray}
By solving for $p_{11}^n, p_{22}^n, \alpha^n$ in terms of $\beta^n$,
we can rewrite these equations so that we have an equation for
$\beta^n$ in terms of $\beta^{n+1}, \beta^{n-1}, \beta^{n+2},
\beta^{n-2}$. This is equivalent to a matrix equation involving a
band-diagonal matrix with five non-zero diagonals. The matrix
equation we must solve is,
\begin{equation}
0 = \mathbf{M}{\underline{\beta}}- E_J \underline{d}/(2\hbar),
\label{eq:fmat1}
\end{equation}
where $\underline{\beta}$ is the vector of the coefficients
$\beta^n$ (with $n$ running from $-\infty$ to $+\infty$),
$\mathbf{M}$ represents the matrix of terms derived from equations
(\ref{eq:fc1})-(\ref{eq:fc4}), and $\underline{d}$ is a vector
representing the Kronecker delta, $\delta_{n,0}$, i.e.\
$\underline{d}(0)=1$ and $\underline{d}(n\neq 0)=0$.

Solving the Fourier series is equivalent to inverting the matrix
$\mathbf{M}$ and the coefficients $\beta^n$ are given by,
\begin{equation}
 \underline{\beta} =   E_J \mathbf{M^{-1}}\underline{d}/(2\hbar) \label{eq:fmat2}.
\end{equation}
The components decay rapidly towards zero for $|n|\gg 1$ and so the
matrix can be truncated and solved numerically with little error. In
practice we found that calculating the coefficients up to $n=\pm 80$
proved more than adequate for the parameters we considered.

\begin{figure}\centering{
\epsfig{file=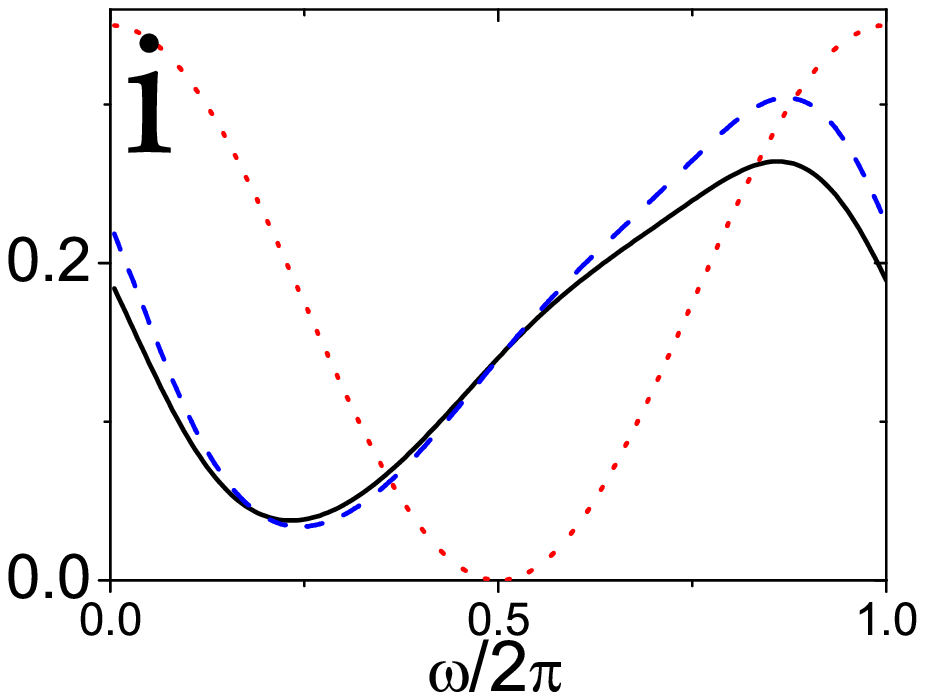, width=6cm}\epsfig{file=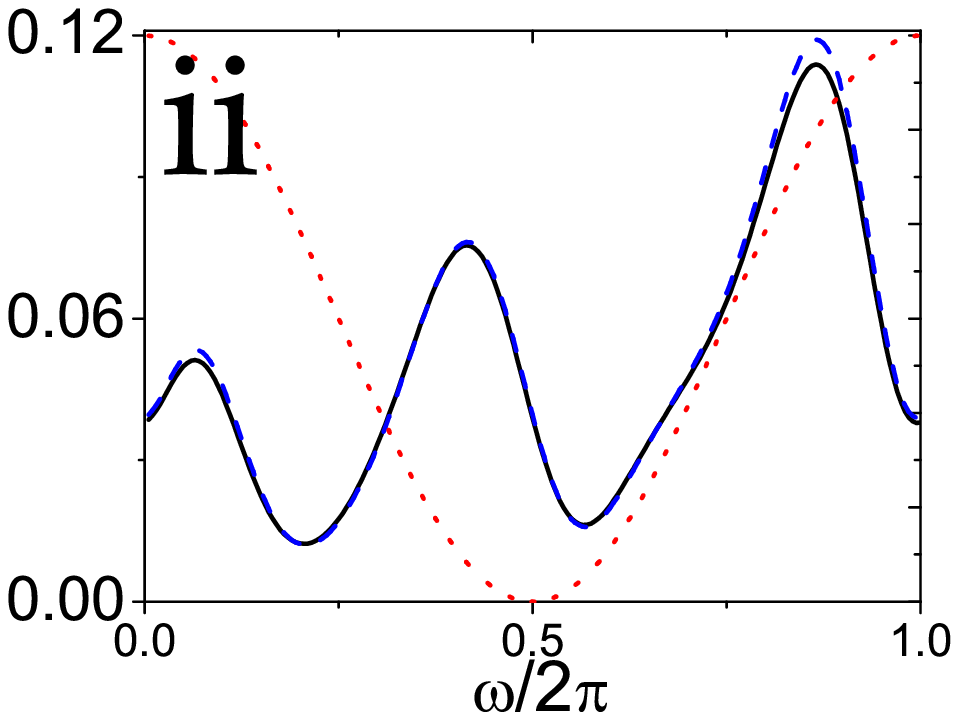,
width=6cm} \epsfig{file=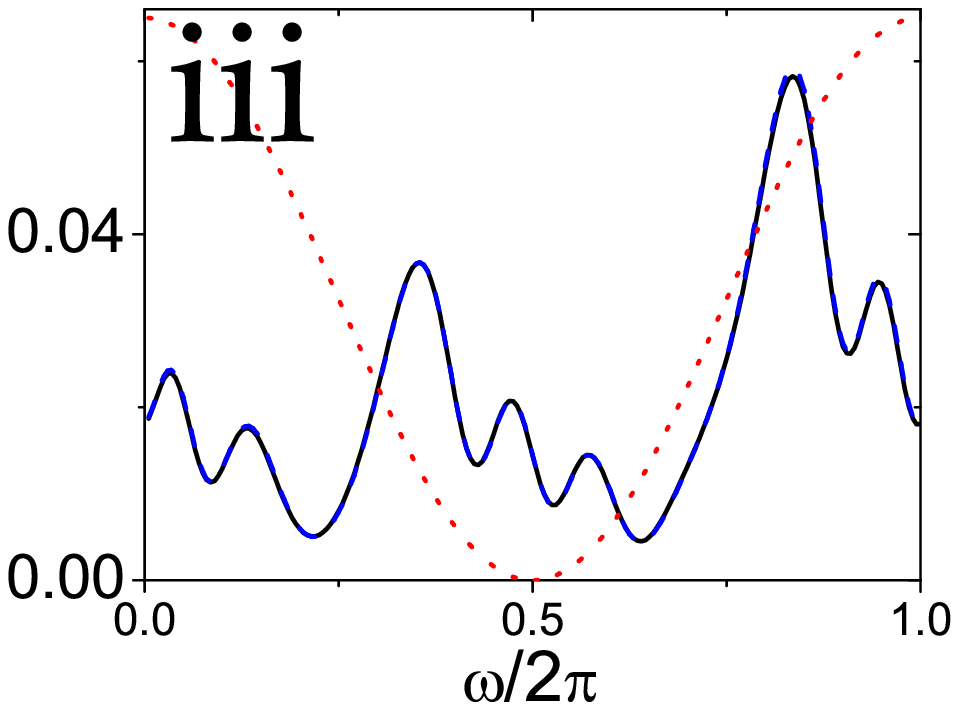, width=6cm}
\epsfig{file=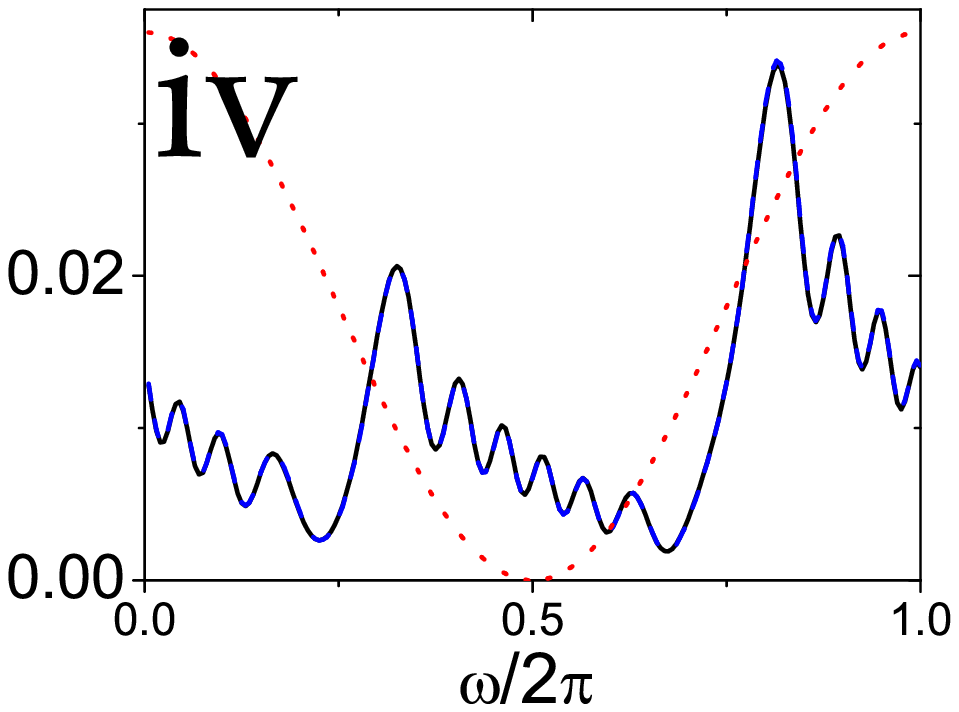, width=6cm}}\caption{Time dependent
oscillations of the SSET charge described by $2p_{22}+p_{11}$,
calculated numerically using the Fourier series solution (solid
curve) and using the analytic approximation described in Sec.\
\ref{sec:analytic} (dashed curve). A harmonic oscillation with the
resonator frequency is plotted (as a dotted curve) for comparison.
The parameters used were $\omega/\Gamma=1, \Delta E/eV_{ds}=-0.1,
\kappa=0.04$, with $A/x_s=$2.4 (i), 9.4 (ii), 21.8 (iii) and 40.7
(iv). } \label{fig:osc}
\end{figure}

Examples of the oscillations in the SSET charge driven by the
resonator are shown in figure \ref{fig:osc} for the case where the
resonator frequency matches the quasiparticle decay rate. We use a
single simplified set of SSET parameters in all numerical
calculations to illustrate the behaviour of our
model\,\cite{SSET2,ria}: $\Gamma=V_{ds}/eR_J$, $R_J=h/e^2$ and
$E_J=hV_{ds}/(16eR_J)$. The SSET-resonator coupling strength is
 parameterized by the dimensionless quantity  $\kappa=m
\omega^2x_s^2/eV_{ds}$.

The charge oscillations in figure \ref{fig:osc} resemble the
behaviour of a periodically kicked damped oscillator, but more
specifically, at larger amplitudes they are very similar to an
oscillator whose frequency is time-dependent. Analogous oscillations
have been seen in the amplitude of radiation within an
optomechanical cavity\,\cite{vahala,MHG}. We can establish the
unforced behaviour of the charge oscillations for a given amplitude
using equations (\ref{mfrphi3})-(\ref{mfrphi6}) and treating the
phase as constant ($\phi=0$). The unforced SSET charge oscillations
have a frequency which increases linearly with the amplitude, $A$.
For sufficiently small $E_J$, we can approximate the frequency by
\begin{equation}
\omega_e\simeq \frac{\Delta E}{\hbar}+\frac{\omega
x_s}{x_q^2}(x_{fp}+A) \label{eq:omegae}
\end{equation}
and the decay rate is $\simeq \Gamma/2$. Thus as the amplitude of
the resonator increases, the number of oscillations in the SSET
charge degrees of freedom during each resonator period increases
accordingly.

\subsection{Effective damping}

Having calculated the response of the SSET charge to the periodic
driving provided by the resonator, we can now calculate how the
resonator, in turn, responds to the dynamics of the SSET. Our
assumption is that the change in the amplitude of the resonator
oscillations is much slower than the oscillations themselves. We can
therefore average the effect of the SSET charge dynamics on the
resonator over a single resonator period, over which time the
amplitude of the resonator motion can be treated as constant.
Integrating equation (\ref{mfrphi1}) over the resonator period
gives,
\begin{eqnarray}
\frac{d \bar{A}}{dt} &=& -\frac{\gamma_{ext}
\bar{A}}{2}-\frac{\omega}{2\pi}\int\limits_t^{t+\frac{2\pi}{\omega}}
dt' \omega x_s({p}_{11}+2{p}_{22})\sin{\omega t'},\nonumber\\
\end{eqnarray}
where the bar on the amplitude, $\bar{A}$, indicates that the
equation is only valid on time scales longer than $2\pi/\omega$. In
terms of the Fourier series, the integration eliminates all of the
terms except those with $n=\pm1$, hence the equation of motion for
$\bar{A}$ can be written as
\begin{eqnarray}
\frac{d \bar{A}}{dt} &=& -\frac{\gamma_{ext} }{2}\bar{A} -x_s\omega
{\rm Im} \left[ {p}^1_{11}(\bar{A})+2{p}^1_{22}(\bar{A})\right].
\label{eq:rbar}
\end{eqnarray}
The second term can be interpreted as an amplitude dependent damping
arising from the interaction with the SSET\,\cite{Usmani,SSET1},
$\gamma_{SSET}(\bar{A}) \bar{A}=2\omega x_s{\rm Im} \left[
{p}^1_{11}(\bar{A})+2{p}^1_{22}(\bar{A})\right]$ as it appears on
the same footing as the external damping rate, $\gamma_{ext}$.
However, when $\gamma_{SSET}(\bar{A})<0$ it means that the charges
transfer energy {\it to} the resonator. The resonator has a constant
amplitude (but not phase) whenever $d \bar{A}/dt=0$, and hence
supports limit cycle solutions whenever this condition is met for
$\bar{A}\neq 0$.

\begin{figure}\centering{
\epsfig{file=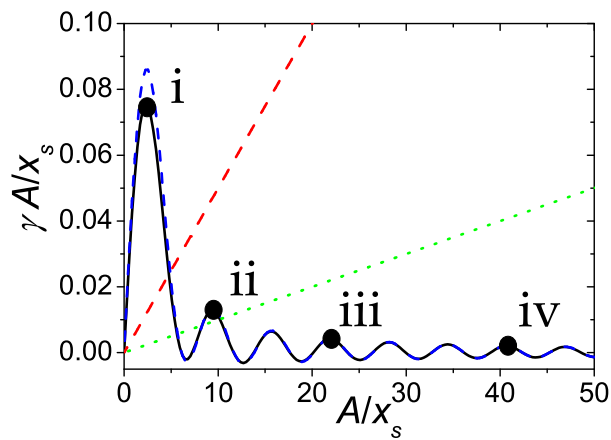, width=8cm}}\caption{Behaviour of the SSET
effective damping, $\gamma_{SSET}$, as a function of the resonator
amplitude, $A/x_s$. The curves show $-\gamma_{SSET}A/x_s$ (in units
where $\Gamma=1$) calculated numerically (full curve) and an
analytic approximation (dashed curve) calculated using equation
(\ref{eq:analytic}). The parameters used are the same as in figure
\ref{fig:osc}. Also plotted are lines indicating the external
damping for $\gamma_{ext}/\Gamma=0.005$ and 0.001 with the latter
being the less steep. Intersections between the curves and one of
the lines indicates the presence of a limit cycle. The points
labeled (i)--(iv) correspond to the oscillations shown in figure
\ref{fig:osc}. } \label{fig:gamofr}
\end{figure}

Figure \ref{fig:gamofr} shows\,\cite{Anote}
$\gamma_{SSET}(\bar{A})\bar{A}$ calculated numerically as a function
of $\bar{A}$, along with various values of $\gamma_{ext}$. Limit
cycle solutions exist for the values of $\bar{A}$ where the curve of
$-\gamma_{SSET}(\bar{A})\bar{A}$ is crossed by a given line
representing $\gamma_{ext}\bar{A}$. The stability of the limit cycle
solutions depends on the gradient of $d\bar{A}/dt$ with respect to
$\bar{A}$ in the usual way, hence a limit cycle is stable where this
gradient is negative\,\cite{strogatz}. Thus we can see by inspection
from figure \ref{fig:gamofr} that for the case where
$\gamma_{ext}/\Gamma=0.001$, there are three possible limit cycle
solutions two of which are stable (the largest and smallest
amplitude ones).

It is clear from figure \ref{fig:gamofr} that it is the oscillations
in $\gamma_{SSET}(\bar{A})$ as a function of $\bar{A}$ that lead to
the existence of more than one limit cycle solution\,\cite{foot} for
sufficiently weak $\gamma_{ext}$. The oscillations in
$\gamma_{SSET}(\bar{A})$ can in turn be understood as arising from
changes in the commensurability of the SSET charge oscillations with
the resonator frequency as the amplitude is increased. This effect
can be seen clearly by comparing the oscillations in figure
\ref{fig:osc} with those in figure \ref{fig:gamofr}. At the first
peak in $-\gamma_{SSET}\bar{A}$, the charge undergoes one
oscillation during the resonator period, a number which increases by
two at each subsequent peak in $-\gamma_{SSET}\bar{A}$.

\subsection{Approximate analytic solution} \label{sec:analytic}

Although the SSET charge is bounded within a narrow range of values,
its oscillations (as shown in figure \ref{fig:osc}) nevertheless
resemble those of a driven harmonic system. Essentially this is
because the parameters chosen are such that the Josephson energy is
not sufficient to allow the charge to saturate over the time scale
of the oscillations. This similarity suggests that it should be
possible to find an approximate solution by reducing equations
(\ref{mfrphi3})-(\ref{mfrphi6}) to an equation for an appropriately
driven harmonic oscillator. We can then follow the approach used in
\,\cite{MHG} where the harmonic dynamics of an optical cavity
coupled to an oscillating mirror was analyzed.

For an uncoupled SSET\,\cite{SSET1,SSET2}, it is straightforward to
show that in the limit where $E_J/\hbar\Gamma\ll 1$ the populations
of the upper charge states, $p_{11}$ and $p_{22}$ always remain much
less than unity. Furthermore, examining the numerical evolution of
the mean field equations we find that this remains true in the
coupled system. This suggests that we can simplify the analysis in
the limit where $E_J/\hbar\Gamma\ll 1$ by neglecting the $p_{11}$
and $p_{22}$ terms in equation (\ref{mfrphi6}). Making this
approximation, and again assuming that the resonator damping is much
slower than the other time scales, we find that equations
(\ref{mfrphi5}) and (\ref{mfrphi6}) reduce to the equations of
motion for a damped harmonic oscillator with a time dependent
frequency term. Using $p_{02}=\alpha+i\beta$, we obtain a single
complex equation of motion,
\begin{eqnarray}
\dot{p}_{02}=\left\{ i \left[\frac{\Delta E}{\hbar}+\frac{\omega
x_s}{x_q^2}(x_{fp}+A\cos{\omega
t})\right] -\frac{\Gamma}{2}\right\}p_{02} + i \frac{E_J}{2\hbar}\nonumber\\
\end{eqnarray}
which can be solved by use of a Fourier series,
$p_{02}=e^{i\theta}\sum e^{i \omega n t} \tilde{p}_{02}^n$ where the
tilde indicates that the Fourier series includes a global phase
shift $\theta(t)=z\sin \omega t$ and $z=Ax_s/x_q^2$. The value of
$\tilde{p}_{02}^n$ can be written in terms of Bessel functions of
the first kind, $\tilde{p}_{02}^n=\widetilde{\psi}^n J_n(-z)$, where
the parameter $\widetilde{\psi}^n$ is given by,
\begin{eqnarray}
\widetilde{\psi}^n=\frac{i E_J/\hbar}{2(i\omega n - i (\Delta
E/\hbar+ \omega x_s x_{fp}/x_q^2) - \Gamma/2) }.
\end{eqnarray}

As before, the damping is calculated from the Fourier coefficients
of $p_{11}$ and $p_{22}$, which in turn can be calculated from
$\beta^n$ using equations (\ref{eq:fc1}) and (\ref{eq:fc2}),
\begin{eqnarray}\label{eq:analytic}
A\gamma_{SSET}= -\omega x_s \frac{E_J}{\hbar}{\rm Im}\left[ \left(
\frac{2}{\Gamma+i \omega}+\frac{\Gamma}{(\Gamma+i\omega)^2}\right)
\beta^1\right].
\end{eqnarray}
In order to make use of this Bessel function solution, we must
convert between Fourier series with and without global phase shifts.
We find,
\begin{eqnarray}\label{eq:bessb}
\beta^n &=& \frac{1}{2i}\sum\limits_{m}\left(
\widetilde{\psi}^{-m}J_{m+n}(z)-(\widetilde{\psi}^{-m})^*J_{m-n}(z)\right)J_m(z).
\end{eqnarray}
This approximate solution can be used to calculate $p_{11}(t),
p_{22}(t)$ (figure \ref{fig:osc}) and hence the effective damping
(figure \ref{fig:gamofr}). It is noticeable in figure
\ref{fig:gamofr} that even for our relatively large choice of
$E_J/\hbar\Gamma\simeq 0.4$, the analytical approximation agrees
well with the numerics at large amplitudes (i.e.\ for $A/x_s\gg 1$).
This is because as the amplitude of the resonator increases, the
driven oscillations which develop in the SSET charge degrees of
freedom become progressively faster and the populations $p_{11}(t)$
and $p_{22}(t)$ become ever smaller as can be seen in figure
\ref{fig:osc}. Using smaller values of $E_J$ leads rapidly to better
agreement at small amplitudes.

\section{Comparison with numerical results}

Much of the usefulness of our analysis of the mean field equations
rests on the degree to which this simplified description of the
SSET-resonator system faithfully reproduces the behaviour seen in
the full numerical solution of the master equation\,\cite{ria}. The
mean field equations allow us to calculate the amplitude of limit
cycles in the resonator dynamics as a function of all the various
parameters of the system, to determine which of them are stable, and
calculate the associated SSET current. Of course the mean field
description does not describe the noise in the SSET-resonator system
and hence it cannot tell us the degree to which a particular limit
cycle may be occupied in regions of parameter space where there is
more than one stable limit cycle (or a stable fixed point solution
coexists with one stable limit cycle).

\subsection{Size of limit cycles}

\begin{figure}[t]
\centering{ \epsfig{file=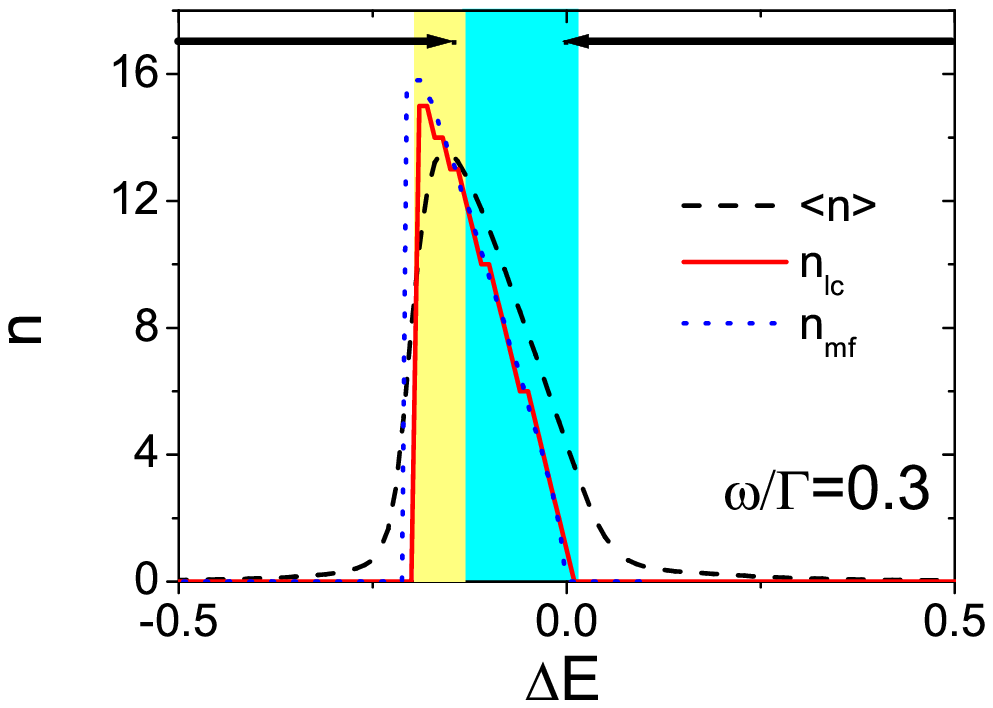,width=7.5cm}
\epsfig{file=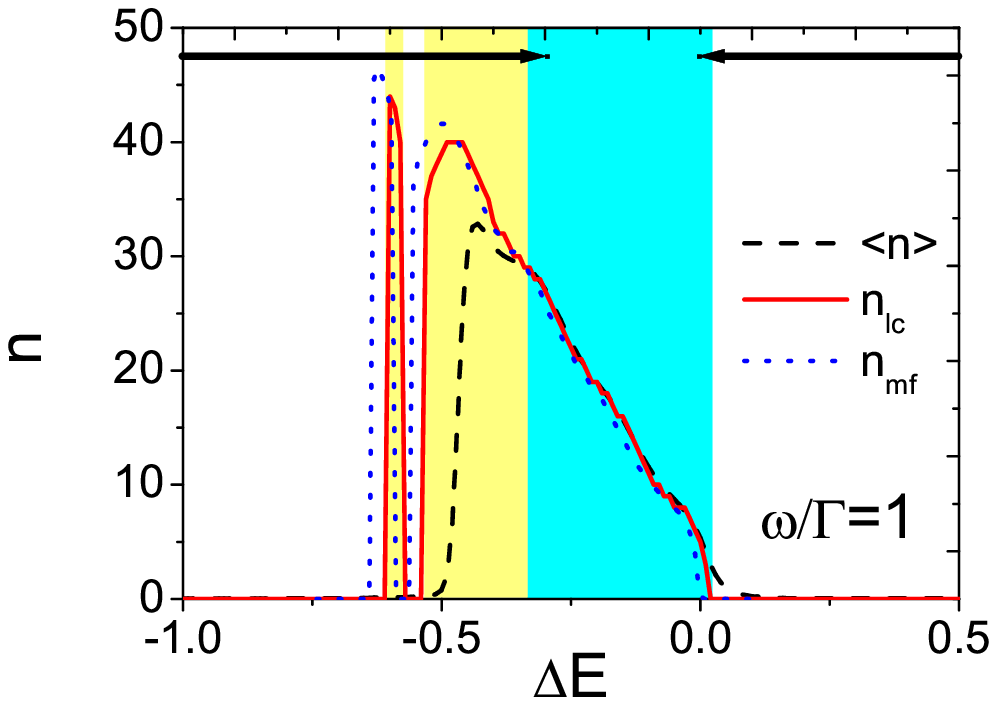,width=7.5cm}\\
\epsfig{file=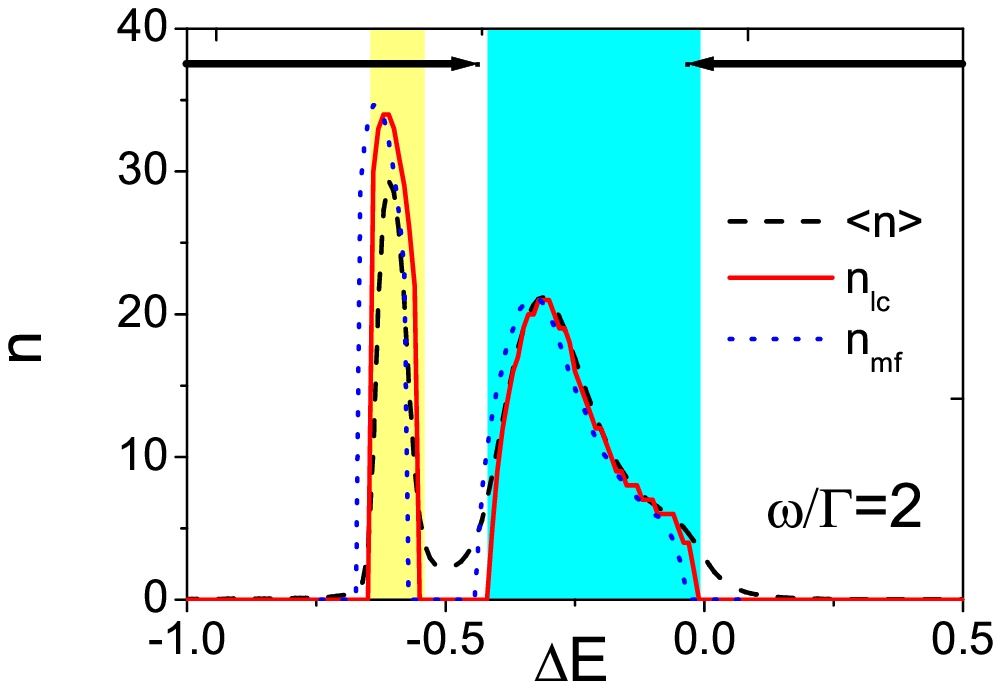,width=7.5cm} \epsfig{file=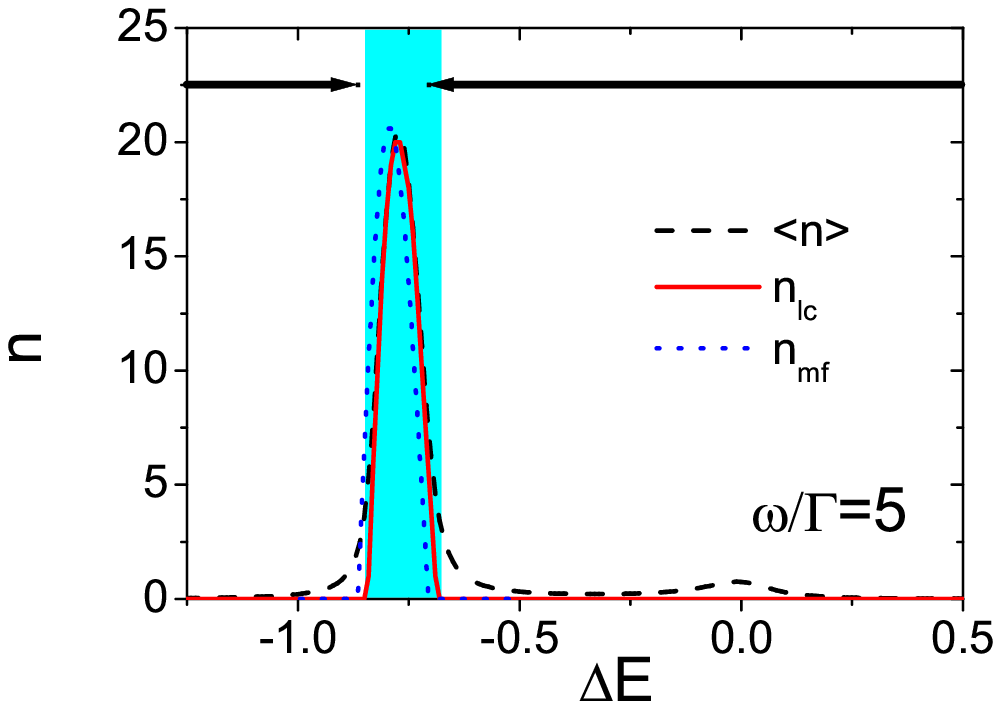,
width=7.5cm}}
 \caption{The size of the limit cycles as a function
of $\Delta E/eV_{ds}$ for a range of values of $\omega/\Gamma$,
with, $\gamma_{ext}/\Gamma=0.002, \kappa=0.01$. The prediction of
the mean field theory, $n_{mf}$ is compared with $n_{lc}$ and
$\langle{n}\rangle$ obtained from the numerical solution of the
master equation. The darker (blue) shaded regions indicate limit
cycle states and the lighter (yellow) shaded region indicates the
regions of bistability, as determined from the master equation. Also
shown for comparison are the regions where a stable fixed point
solution exists calculated using the fixed point analysis described
in Appendix B (arrows).} \label{fig:deltae}
\end{figure}

We begin by considering sufficiently weak couplings (for a given
value of $\gamma_{ext}$) that the resonator is limited to at most a
single limit cycle state\,\cite{ria} and examine when such states
develop and their sizes as a function of the detuning from
resonance, $\Delta E$, and the relative speed of the resonator,
$\omega/\Gamma$. In figure \ref{fig:deltae} the size of the stable
limit cycles calculated using the mean field equations [i.e., using
equation (\ref{eq:rbar})] is compared with the numerical solution of
the full master equation as a function of $\Delta E$ for various
resonator frequencies.

The numerical solution of the master equation gives us the full
steady-state density matrix of the system, $\rho_{ss}$, in terms of
which the probability of finding the resonator in a given Fock
state, $|n\rangle$, is simply $P(n)={\rm Tr}[|n\rangle\langle n|
\rho_{ss}]$. Because of the ensemble average and the presence of
phase noise, limit cycles do not appear as periodic features in the
dynamics of the density matrix, but they can be identified as peaks
in the distribution $P(n)$ above the ground state energy (i.e.\
$n>0$). [These peaks typically correspond to individual rings in the
associated Wigner function representation of the density
matrix\,\cite{ria}.] We have used two quantities from the $P(n)$
distribution to compare with the mean field results: the average
number of resonator quanta, $\langle{n}\rangle=\sum_n nP(n)$, and
the $n$-value of the peak in the distribution, which we define as
$n_{lc}$.
 In order to compare these values with the stable limit cycle amplitudes calculated
 in the mean field theory, we express the latter in terms of resonator
  quanta, $n_{mf}=A^2(m\omega/2\hbar)$.

 It is clear from  figure \ref{fig:deltae}  that the mean field equations prove
  to be a rather good predictor of the locations and
sizes of the limit cycles in the full system dynamics as can be seen
by the relatively good agreement between $n_{mf}$ and $n_{lc}$. It
is interesting to note that a comparison of the sizes of the limit
cycles with $\langle{n}\rangle$ works well in regions where the
resonator energy distribution $P(n)$ is relatively concentrated.
Thus $n_{mf}$ is quite close to $\langle{n}\rangle$ when there is a
single, large, limit cycle state present, but not in the vicinity of
the transitions where the limit cycles begin to form or in regions
where a limit cycle state coexists with a stable fixed point
(bistable regions).

\begin{figure}\centering{
\epsfig{file=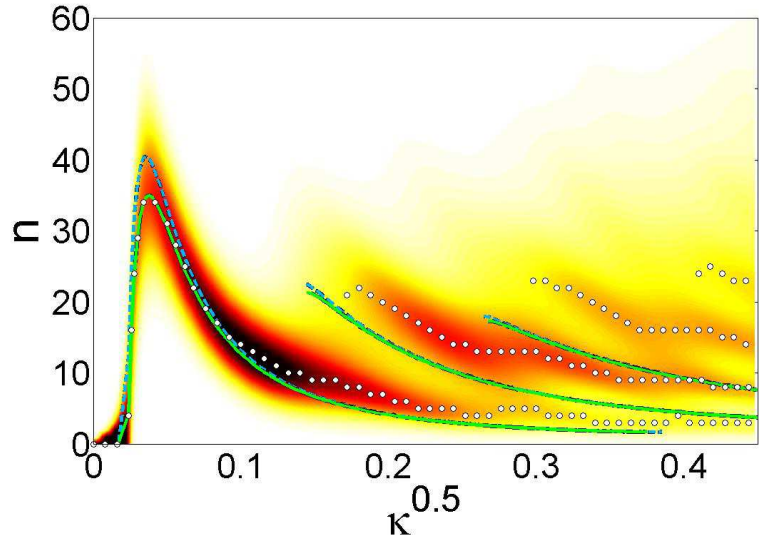, width=10cm}}\caption{The size of the
limit cycles as a function of $\kappa^{1/2}$ calculated numerically
and using the mean field equations, with $\omega/\Gamma=1$,
$\gamma_{ex}/\Gamma=0.001$, $\Delta E/eV_{ds} = -0.1$. The solid
green line denotes the mean field solution, and the dotted blue
curve indicates the size of the limit cycles calculated using the
approximate Bessel function solution. The background shading shows
the numerically calculated steady state distribution of the
resonator, $P(n)$, with dark red indicating highest probability. The
white dots indicate the peaks in $P(n)$.} \label{fig:kapsweep}
\end{figure}

Despite the generally good agreement between $n_{mp}$ and $n_{lc}$,
figure \ref{fig:deltae} reveals that there are small differences in
the locations of the values of $\Delta E$ at which the limit cycles
appear (or disappear) predicted by the mean field equations and the
master equation. This can partly explained by the fact that our
analysis is most appropriate when the limit cycles are large on a
scale set by $x_s$ (as discussed in Sec.\ \ref{sec:radial}).
However, the differences in onset points for the limit cycles exist
not just in cases where the limit cycles grow continuously from zero
(around $\Delta E=0$), but also when they emerge with a relatively
large radius. An interesting possible explanation for these
observations is that the exact location of the dynamical transitions
may in fact depend quite sensitively on the noise in the system---an
element which is of course missing in our mean field
analysis\,\cite{mabuchi,bc}.

 We now turn to compare the mean field analysis with numerical
predictions for a given detuning, $\Delta E$, as the SSET-resonator
coupling is increased. For resonator frequencies of the order of the
quasiparticle decay rate, numerical solution of the master equation
showed that increasing the coupling could lead to a sequence of
transitions marked by the appearance of increasing numbers of peaks
in the steady-state distribution $P(n)$\,\cite{ria}.

 In figure \ref{fig:kapsweep}
we plot the size of the stable limit cycles calculated both
numerically using the Fourier series solution of the mean field
equations and using the Bessel function expression [equation
(\ref{eq:analytic})] as a function of $\kappa^{1/2}$ for
$\omega/\Gamma=1$. The predictions of the mean field equations are
compared with the locations of the peaks in the numerically
calculated resonator distribution, $P(n)$. The mean field
calculation shows good qualitative agreement with the full numerics,
showing a series of bifurcations and multiple limit cycles as the
coupling is increased. For weaker coupling, $\kappa^{1/2}\lesssim
0.1$, the mean field calculation accurately predicts the size of the
limit cycle. Notice, however, that the Bessel function approximation
for the limit cycle differs appreciably from the full mean field for
the first limit cycle, but otherwise matches the mean field result
closely. As before, reducing the value of $E_J$ improves the
accuracy of the Bessel function approximation.

The appearance of successive stable limit cycle solutions as the
coupling is increased in figure \ref{fig:kapsweep} is readily
understood in terms of the analysis of the charge oscillations given
in Sec.\ \ref{sec:fourier} and the associated oscillations in
$\gamma_{SSET}(A)$ (illustrated in figure \ref{fig:gamofr}). From
equation (\ref{eq:omegae}) we see that increasing the SSET-resonator
coupling (i.e.\ increasing $x_s$) increases the frequency of the
charge oscillations thus changing their commensurability with the
mechanical period. This
 effectively compresses the oscillations of
$\gamma_{SSET}(A)$ as a function of $A$ (i.e.\ they occur with a
progressively smaller period measured in terms of amplitude). Thus
for fixed $\gamma_{ext}$, increasing the coupling means that more
and more stable limit cycle solutions occur and those already
present move to smaller and smaller sizes.

\subsection{SSET Current}

It is also possible to calculate the current through the SSET using
the mean field equations. The current is generated by quasiparticle
tunnelling out of the states $|1\rangle$ and $|2\rangle$, which
leads to a time dependent tunnel current across the right  junction,
$I(t)=\Gamma(p_{11}(t)+p_{22}(t))$. Therefore, when the resonator is
in a limit cycle state of a particular amplitude, the corresponding
oscillations in $p_{11}(t)$ and $p_{22}(t)$ [see figure
\ref{fig:osc}] will be passed on to the tunnel current. The average
current (defined as either an average over one period of mechanical
oscillation, or over an ensemble of systems) of course will not
reflect these oscillations, but it will depend on the amplitude of
the resonator's motion. For a given resonator amplitude, the
corresponding current is given by the Fourier coefficients,
$\langle{I}\rangle/e=\Gamma(p^0_{11}({A})+p^0_{22}({A}))$.

The average current calculated using the mean field equations only
agrees well with that calculated using the full master equation when
the steady state of the latter predicts a narrow width to the
distribution $P(n)$. However, we expect that the current calculated
within the mean field picture will be useful in understanding the
current noise in the SSET\,\cite{cnoise}. For example, we might
expect signatures of the different frequencies of charge
oscillations corresponding to different resonator states to appear
in the current noise spectrum.

\section{Practical Implications}

The mean field analysis can be used to provide simple estimates of
the kinds of dynamical instabilities that may be seen in a
particular experiment. Recent experiments on a 22MHz nanomechanical
resonator coupled to a SSET were carried out\,\cite{SSET-expt} in
the regime where $\omega/\Gamma\ll 1$. Previous
calculations\,\cite{SSET1,bc,SSET2} showed that it should be
possible to see the transition from the fixed point to a limit cycle
state in such systems and a similar conclusion is reached using the
mean field equations. Indeed some evidence for instabilities in the
resonator motion was found in these experiments, although the
primary focus of the work was on the stable
regimes\,\cite{SSET-expt,Naik}.

\begin{figure}\centering{
\epsfig{file=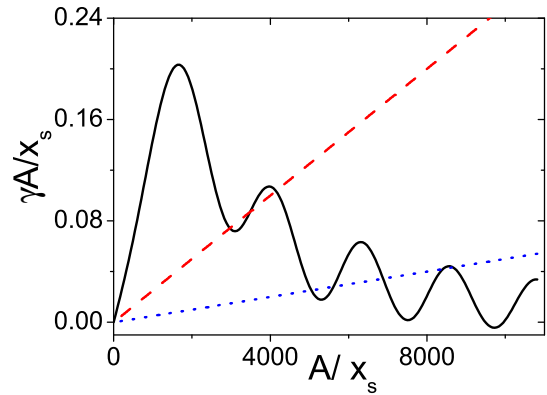, width=8cm}}\caption{Behaviour of the
SSET effective damping, $\gamma_{SSET}$, as a function of the
resonator amplitude, $A/x_s$. The parameters used are as described
in the text below. The curve shows $-\gamma_{SSET}A/x_s$, while the
straight lines indicate the external damping for
$\gamma_{ext}/\Gamma=2.25 \times 10^{-5}$ and $5\times 10^{-6}$ with
the latter being the less steep.} \label{fig:paramrsweep}
\end{figure}

It is natural to use the mean field approach to estimate whether the
regime where the resonator dynamics involves more than one limit
cycle solution is accessible experimentally. For the nanomechanical
resonator and SSET parameters in the experiments of Naik et
al.\,\cite{SSET-expt} we find that the wide separation of
time-scales, $\omega/\Gamma\ll 1$, means that the values of the
SSET-resonator coupling and $\gamma_{ext}$ necessary to reach this
regime are well beyond currently achievable values. However, the
interaction between the SSET and resonator is much stronger when
$\omega/ \Gamma \sim 1$, which suggests that the observation of more
than one limit cycle may be possible when the resonator and
quasiparticle decay timescales are more closely matched.

Although nanomechanical resonators with frequencies $\sim 1$GHz have
been produced,\,\cite{GHZ} integration with SET electronics has only
been achieved for resonators up to $\sim 100$MHz\,\cite{SET-expt}.
However, it is possible to reduce the quasiparticle tunnelling rate,
$\Gamma$, substantially by increasing the resistance of the relevant
tunnel junction and rates $\leq 2\times 10^8$s$^{-1}$ have been
demonstrated\,\cite{Nakamura}. Therefore, as an example, we consider
a nanomechanical resonator with fundamental frequency
$\omega/2\pi=100$MHz coupled to a SSET in which all the charge
processes are rather slower than usual. Quasiparticle tunnelling in
the SSET is assumed to occur across a junction with very high
resistance, 5M$\Omega$, and a Josephson energy at the other junction
which is tuned (e.g. using the method employed by Nakamura et
al.\,\cite{Nakamura}) to be $E_J=1\times10^{-3} eV_{ds}$. For the
other parameters of the system we choose values which are within the
general range explored in recent
experiments\,\cite{SSET-expt,SET-expt,SET-expt2}. The other SSET
parameters we assume are $E_C=175\mu$eV, $eV_{ds}=700\mu$eV, and
$\Delta =200\mu$eV. For the resonator we assume a mass
$m=6.8\times10^{-16}$kg, with a SSET-resonator separation $d=100$nm,
a SSET-resonator capacitance $C_{g}=34$aF and coupling voltage
$V_{g}=1$V. We note that this choice of parameters takes us outside
the limit $E_J/\hbar\Gamma \ll 1$ and we must rely on the numerical
solution to the full Fourier series (rather than use the Bessel
function approximation).

In figure \ref{fig:paramrsweep} we plot the effective SSET damping
as a function of resonator amplitude for a detuning
 $\Delta E= -1\times10^{-3} eV_{ds}$. From figure 6, we see that we first get two stable
limit cycles around $\gamma_{ext}/\Gamma\simeq 2.5\times 10^{-4}$,
corresponding to a resonator quality factor of $2.9\times 10^{4}$,
which should be accessible experimentally. The observation of
multiple limit cycles of course requires that the fluctuations in
resonator amplitude are not so large as to wash out the difference
in amplitudes between the cycles. Although a full calculation of the
noise in the system is beyond the mean field theory as presented, we
can make an estimate of the length scale of the fluctuations due to
external thermal noise which, for the parameters in figure
\ref{fig:paramrsweep} and a temperature of 30mK we find to be about
$\delta A \simeq 136 x_s$ i.e. about 10\% of the separation in
amplitude of the limit cycles. This suggests that multistability
will occur within a region accessible by experiment though the
conditions required to see it are quite demanding.

\section{Conclusions}

The mean field analysis presented here provides a simplified
description of the SSET-resonator system. In particular, the mean
field approach provides a good description of the dynamical
instabilities which the resonator can undergo. For relatively weak
SSET-resonator couplings the mean field equations describe the onset
and size of the first limit cycle state quite accurately for a wide
range of resonator frequencies. As the coupling is increased, the
mean field equations predict the emergence of a succession of limit
cycle states of different sizes. Numerical calculations based on the
full master equation reveal that although the mean field equations
become progressively less accurate as the coupling is increased they
still give a good qualitative description of the dynamical
behaviour. Furthermore, the mean field analysis allows us to relate
the appearance of further limit cycles to changes in the
commensurability of oscillations in the SSET charge with the period
of the resonator.

\ack

This work was supported by the EPSRC under grants GR/S42415/01 and
EP/C540182/1.

\appendix
\section{SSET-resonator master equation}

In this appendix we outline some of the steps in the derivation of
the master equation, equation (\ref{master}) and the approximations
made. Essentially the same master equation was introduced in Ref.\
\cite{SSET2} and used again in \cite{ria}, but details of its
derivation have not yet been given.

In essence the calculation is a straightforward generalization of
the treatment of the JQP cycle given in Ref.\ \cite{Choi} to include
coupling to a resonator. The Hamiltonian for the SSET-resonator
system can be written as the sum of several parts,
\begin{equation}
H=H_{\rm C}+H_{\rm J}+H_{\rm R}+H_{\rm qp}+H_{\rm T}
\end{equation}
which correspond to the charging energy of the SSET, $H_{\rm C}$,
coherent Josephson coupling between the left  lead and the island,
$H_{\rm J}$, the energy of the resonator, $H_{\rm R}$, the energy of
the quasiparticles in the leads and island of the SSET, $H_{\rm qp}$
and the quasiparticle tunnelling Hamiltonian which couples the leads
and the island, $H_{\rm T}$.

As discussed in Sec.\ \ref{sec:me}, for relatively small resonator
displacements the resonator-charge coupling can be linearized and
hence the charging energy of the SSET written
as\,\cite{footnote_freq}
\begin{eqnarray}
H_{\rm C}&=&\sum_{N,n}\left[E_c\left(N^2-2NN_g\right)-neV_{ds}
+m\omega^2x_sxN\right]|N \rangle\langle N|\otimes|n \rangle\langle
n| \label{charging}
\end{eqnarray}
where $E_c=e^2/(2C_J+C_g)$, $N_g=(C_gV_g+C_JV_{ds})/e$ and
$x_s=2E_cC_gV_g/(em\omega^2d)$. The macroscopic charge variables $N$
and $n$, which correspond to charge states $|N\rangle$ and $|
n\rangle$, are the number of excess electronic charges on the SSET
island and the number of electrons which have tunnelled {\it off}
the island via the right hand junction respectively. The Hamiltonian
of the resonator is simply that of a harmonic oscillator,
\begin{equation}
H_{\rm R}=\frac{p^2}{2m}+\frac{1}{2}m\omega^2x^2.
\end{equation}
For the bias configuration we consider, Josephson tunnelling is only
important between the left lead and the island as tunnelling between
the right  lead and the island involves energy differences $\sim
2eV_{ds}\gg E_J$ because of the bias voltage applied (we neglect the
possibility of the resonator having large enough energy to assist in
overcoming this barrier). Thus we include just the coherent
tunnelling between the island the left  lead (i.e.\ coherent
tunnelling changes $N$ but not $n$),
\begin{equation}
H_{\rm J}=-\sum_N \frac{E_J}{2}\left( |N\rangle\langle
N+2|+|N+2\rangle\langle N|\right).
\end{equation}

The quasiparticle Hamiltonian is given by\,\cite{Choi}
\begin{equation}
H_{\rm qp}=\sum_{\alpha=L,R,I}\sum_{k,\sigma}\epsilon_{k
\alpha}c^{\dagger}_{k\alpha\sigma}c_{k\alpha\sigma},
\end{equation}
where the sum $\alpha$ runs over the three pieces of superconductor
(left lead $L$, right  lead $R$, and island $I$) and the operator
$c^\dagger_{k\sigma L}$ creates a quasiparticle in the left lead
with energy $\epsilon_{kL}$, momentum $k$  and spin state $\sigma$.
Quasiparticle tunnelling between the island and the leads is
described by the Hamiltonian,
\begin{equation}
H_{\rm T}=\sum_{j=L,R}\left({\rm e}^{- i\phi_j/2}X_j+{\rm e}^{+
i\phi_j/2}X^{\dagger}_j\right)
\end{equation}
where
\begin{equation}
X_j=\sum_{k,q,\sigma}T_{kq}c^{\dagger}_{kj\sigma}c_{qI\sigma}
 \end{equation}
 creates a quasiparticle in state $k$ in lead $j$ and destroys one
 in state $q$ in the island with tunnelling amplitude $T_{kq}$. The operators ${\rm e}^{\pm
i\phi_j/2}$ describe the associated change in the macroscopic charge
variables, in terms of which they are written
\begin{eqnarray}
{\rm e}^{\mp i\phi_R/2}&=&\sum_{N,n}|n\pm 1\rangle\langle
n|\otimes|N\mp
1\rangle\langle N|\\
{\rm e}^{\mp i\phi_L/2}&=&\sum_{N}|N\mp 1\rangle\langle N|.
\end{eqnarray}

The master equation for the macroscopic island charge $N$ and
resonator is obtained by tracing out the quasiparticle degrees of
freedom and the count variable $n$. The procedure is very similar to
that used to derive the analogous master equation for a normal-state
SET\,\cite{dar}. The derivation uses the standard Born-Markov
approach\,\cite{WM} which involves the assumptions that the
tunnelling Hamiltonian $H_{\rm T}$ is a weak perturbation on the
quasiparticle Hamiltonian and that the quasiparticles relax back to
their unperturbed distributions after tunnelling faster than any
other time-scale in the problem.

Further simplification is achieved by limiting the analysis to
include only those states which are accessible to the system in the
zero temperature limit. The charging energy of the SSET island,
$E_c$ is typically much larger than $E_J$, hence the Josephson
tunnelling can be neglected for all states except the two charge
states which are selected at the JQP resonance (by
tuning\,\cite{footnote_extra} $N_g$, see equation \ref{charging}) to
be almost degenerate. For simplicity we consider states $|0\rangle$
and $|2\rangle$, which differ by one Cooper pair, corresponding to
resonance (for a fixed gate) at $N_g=1$. For quasiparticle decay to
occur, the energy gained when a particle tunnels from the island to
a lead must be $\ge 2\Delta$, where $\Delta$ is the superconducting
gap\,\cite{SSET2}. At the JQP resonance the voltage $V_{ds}$ is
chosen so that only two processes are allowed: tunnelling from the
island into the right lead between states $|2\rangle$ and $|
1\rangle$, and between $|1\rangle$ and $|0\rangle$. The displacement
of the resonator produces changes in the electrostatic energy
differences involved in quasiparticle tunnelling which are assumed
to be small with respect to the values for a fixed gate. This is the
reason for our choice of a dimensionless of coupling,
$\kappa=m\omega^2x_s^2/eV_{ds}\ll1$, which measures the energy
associated with coupling to the resonator in terms of the typical
energy scale associated with the unperturbed quasiparticle
tunnelling rates. The question of when other charge states become
accessible (i.e.\ for very strong SSET-resonator couplings or high
enough temperatures), and what effect they have on the dynamics is
an interesting one, but we do not consider it here.

Taking all these factors into account, we obtain a master equation
confined to the space of the three charge states $|2\rangle$,
$|1\rangle$ and $|0\rangle$,
\begin{eqnarray}
\fl \dot{\rho}=-\frac{i}{\hbar}[H_{\rm co},\rho]\label{lmaster}\\
\fl-\frac{\Gamma( E_{2,1})}{2}\left[\left\{|2\rangle\langle
2|,\rho\right\}_+-|1\rangle\langle 2|\rho(|1\rangle\langle
0|+|2\rangle\langle 1|)-(|0\rangle\langle 1|+|1\rangle\langle
2|)\rho|2\rangle\langle 1|\right] \nonumber\\
\fl-\frac{m\omega^2x_s\Gamma'(E_{2,1})}{2}\left[\left\{x|2\rangle\langle
2|,\rho\right\}_+-x|1\rangle\langle 2|\rho(|1\rangle\langle
0|+|2\rangle\langle 1|)-(|0\rangle\langle 1|+|1\rangle\langle
2|)\rho|2\rangle\langle 1|x\right]\nonumber\\
\fl-\frac{\Gamma( E_{1,0})}{2}\left[\left\{|1\rangle\langle
1|,\rho\right\}_+-|0\rangle\langle 1|\rho(|2\rangle\langle
1|+|1\rangle\langle 0|)-(|0\rangle\langle 1|+|1\rangle\langle
2|)\rho|1\rangle\langle 0|\right] \nonumber \\
\fl-\frac{m\omega^2x_s\Gamma'(
E_{1,0})}{2}\left[\left\{x|1\rangle\langle
1|,\rho\right\}_+-x|0\rangle\langle 1|\rho(|2\rangle\langle
1|+|1\rangle\langle 0|)-(|0\rangle\langle 1|+|1\rangle\langle
2|)\rho|1\rangle\langle 0|x\right],\nonumber
\end{eqnarray}
where $\Gamma(E)$ is the quasiparticle tunnel rate for energy $E$
(this is the energy gained by the quasiparticle when it tunnels from
island to lead) with the relevant energies for quasiparticle
tunnelling from $|2\rangle$ to $|1\rangle$ and from $|1\rangle$ to
$|0\rangle$ given by\,\cite{SSET2},
\begin{eqnarray}
E_{2,1}&=&eV_{ds}+2E_c(3/2-N_g)\\
E_{1,0}&=&eV_{ds}+2E_c(1/2-N_g),
\end{eqnarray}
respectively. We have assumed that the position dependent changes in
the energy differences are small with respect to the typical energy
scale $eV_{ds}$, so that we can expand the tunnelling rates to first
order\,\cite{SSET2},
\begin{equation}
\Gamma(E+m\omega^2x_sx)\simeq \Gamma(E)+m\omega^2x_sx\Gamma'(E)
\end{equation}
where $\Gamma'(E)=d\Gamma(E)/dE$. The Hamiltonian in equation
(\ref{lmaster}) takes the simplified form
\begin{eqnarray}
H_{\rm co}&=&\Delta E|2\rangle\langle 2|+E_1|1\rangle\langle
1|-\frac{E_J}{2}\left(|2\rangle\langle0|+|0\rangle\langle 2|\right)
\nonumber\\&& +m\omega^2 x_sx\left[|1\rangle\langle
1|+2|2\rangle\langle 2|\right]+H_{\rm R},
\end{eqnarray}
where
\begin{eqnarray}
\Delta E&=&4E_c(1-N_g)\\
 E_1&=&4E_c(1-2N_g).
\end{eqnarray}

The master equation we have obtained [equation (\ref{lmaster})] can
be analyzed numerically quite easily. However, it is useful to make
further simplifications which make it easier to identify the
essential physics of the system. The off-diagonal elements of the
density matrix $\langle 0|\rho|1\rangle$, $\langle 1|\rho|2\rangle$
(together with their complex conjugates) decouple from the dynamics
of the other parts of the density matrix and can be dropped,
together with the $E_1$ term in $H_{\rm co}$, as they play no role
in the resonator dynamics. We also assume that the difference
between the two quasiparticle decay rates and their variations with
bias point can be neglected. Whilst the difference between these
rates (and their dependence on the bias point) may be important in
the analysis of a given experimental system, it is not essential to
our theoretical analysis which seeks to describe the basic features
of the system in the simplest possible way.

Finally,
 we also drop the position dependent parts of
the quasiparticle tunnelling terms. All the important dynamics
arises from the position dependence in the coherent part of the
master equation which modulates the detuning from resonance $\Delta
E$ (which can be arbitrarily small). In contrast, the position
dependent coupling which appears in the quasiparticle decay terms is
expected in practice to be much weaker than the coherent position
coupling and provide only a small modulation of the quasiparticle
rates\,\cite{SSET1,SSET2,bc}. Although numerically (within the
master equation formalism) we find that including the position
dependence of the quasiparticle rates can eventually lead to quite
large changes in the steady state distribution function for
sufficiently large $\kappa$, the general pattern and range of
resonator behaviours (including the existence of regions where the
resonator state is number-squeezed) remains essentially the same.
Furthermore, corrections arising from the position dependence of the
quasiparticle rates become progressively smaller as $R_J/(h/e^2)$ is
increased\,\cite{SSET1,bc}. Note, however, that the parameters used
in the main text (and in \,\cite{ria}) are chosen to best illustrate
the behaviour of our simplified model and we have not attempted to
in addition minimize the corrections that would arise if the
position dependence of the quasiparticle rates were included.

\section{Stability analysis of the fixed point}

In this appendix we describe the analysis of the fixed point
solutions of the mean field equations [equations
(\ref{mf1})-(\ref{mf5})]. The fixed point solution, $\beta_{fp}$, is
found from the solution of the following cubic equation,
\begin{eqnarray}
18\left(\frac{E_J}{\hbar\Gamma}\right)^2\left(\frac{x_s}{x_q}\right)^4\frac{\omega^2}{\Gamma}\beta_{fp}^3
+12\frac{E_J\Delta
E}{(\hbar\Gamma)^2}\left(\frac{x_s}{x_q}\right)^2\omega\beta_{fp}^2 \label{B1}\\
+\left(2\frac{\Delta
E^2}{\hbar^2\Gamma}+\frac{3E_J^2}{2\hbar^2\Gamma}+\frac{\Gamma}{2}\right)\beta_{fp}+\frac{E_J}{2\hbar}&=&0,
\nonumber
\end{eqnarray}
in terms of which the fixed point values of the other dynamical
variables are readily obtained. In particular, the fixed point
resonator position is given by,
\begin{equation}
x_{fp}=\frac{3E_J}{\hbar\Gamma}x_s\beta_{fp}.
\end{equation}
It is obvious from equation (\ref{B1}) that there is only a single
solution for $\beta$ in the limit $\kappa\rightarrow 0$. We have
checked numerically that there remains only one physically
acceptable solution over the range of parameters studied here.

In order to establish the stability of the fixed point solution we
need to calculate the Jacobian matrix of the mean field equations at
the fixed point\,\cite{strogatz}. The stability of the system is
then determined by the eigenvalues of this matrix, $\lambda$, which
are solutions of the characteristic equation which can be written as
\begin{equation}
\lambda^6+a_5\lambda^5+a_4\lambda^4+a_3\lambda^3+a_2\lambda^2+a_1\lambda^1+a_0=0
\end{equation}
The coefficients $a_0$, $a_1$, etc.\ are functions of the various
parameters of the system determined by taking the determinant of the
appropriate Jacobian. Whilst it is possible to determine the
stability of the system by simply calculating all the eigenvalues
numerically it turns out that we can establish the stability of the
fixed point solution using a somewhat simpler approach.

 Assuming we start from a stable region
and slowly change one of the parameters of the system, a limit cycle
of the resonator begins to form when a pair of complex eigenvalues
cross the real axis (Hopf
bifurcation)\,\cite{strogatz,mabuchi,ffun}. The locations of these
transitions can be determined by using the fact that at these points
the characteristic equation must support a solution of the form
$\lambda=i\nu$ (with $\nu$ a positive real number). This requirement
can be re-expressed in terms of the condition on the
corefficients\,\cite{mabuchi,ffun}
\begin{equation}
\fl
f=a_5\left(\frac{a_1^2+a_5(a_0a_3-a_1a_2)}{a_5(a_0a_5-a_1a_4)+a_1a_3}\right)^2
-a_3\left(\frac{a_1^2+a_5(a_0a_3-a_1a_2)}{a_5(a_0a_5-a_1a_4)+a_1a_3}\right)+a_1=0
\end{equation}
The regions of stability marked by arrows in figure \ref{fig:deltae}
were determined by evaluating $f$ for each set of parameters. In
stable regions $f<0$, changing sign at the Hopf bifurcations. Notice
that our analysis is centred on the fixed point solution; it tells
us nothing about the possible coexistence of stable fixed point and
limit cycle solutions and hence does not describe the regions of
bistability in figure \ref{fig:deltae}.

\section*{References}

% ----------------------------------------------------------------
%\bibliographystyle{amsplain}
%\bibliography{}

\begin{thebibliography}{99}
\bibitem{footnote_3} The energy difference
between Cooper pair states is renormalized slightly by the
interaction with the resonator leading to a small shift in the
center of the resonance. However, we have used the bare value,
$\Delta E=0$, in discussions and figures as the renormalization is
minimal for weak coupling.
\bibitem{Anote} This plot is essentially a generalization of figure 4 in
Ref.\ \cite{SSET1}, though we have used a somewhat different
parameter regime and a different method.
\bibitem{foot}Note that for the SSET parameters we used, the oscillations in
$\gamma_{SSET}$ were found to be largest for $\omega\simeq
\Gamma/2$. This is consistent with the observation made in Ref.\
\cite{ria} (based on numerical solution of the master equation for a
range of resonator frequencies) that the electro-mechanical coupling
strengths required to produce successive transitions are lowest when
$\omega\sim \Gamma$.
\bibitem{sccase} Note that for a superconducting resonator coupled to Cooper pair
box much higher frequencies ($\geq 5$GHz) have been
demonstrated\cite{SCR}.
\bibitem{footnote_freq} This expression neglects a term which
gives rise to an additional gate-voltage dependent frequency shift
in the resonator.
\bibitem{footnote_extra} In experiment fine tuning of the bias point with respect to the JQP resonance
can be achieved independently of the SSET-resonator coupling using
another fixed gate (see Ref.\ \cite{SSET-expt}).
\end{thebibliography}

\end{document}